\documentclass[a4paper,12pt]{article}
\usepackage{epsfig}

\def\dofigs#1#2#3{\centerline{\epsfxsize=#1\epsfig{file=#2, width=7.5cm, 
height=7.5cm, angle=-90}%
\hfil\epsfxsize=#1\epsfig{file=#3,  width=7.5cm, height=7.5cm, angle=-90}}}

\newcommand{\be}{\begin{equation}}
\newcommand{\ee}{\end{equation}}

\newcommand{\bea}{\begin{eqnarray}}
\newcommand{\eea}{\end{eqnarray}}

\newcommand{\gsim}{\ \rlap{\raise 2pt\hbox{$>$}}{\lower 2pt \hbox{$\sim$}}\ }
\newcommand{\lsim}{\ \rlap{\raise 2pt\hbox{$<$}}{\lower 2pt \hbox{$\sim$}}\ }

\newcommand{\amu}{a_{\mu}}
\newcommand{\asusy}{a_{\mu}^{{ \rm SUSY}}}

\def\tgb{\tan{\beta}}

\def\bsg{$b\to s\gamma$~}

\begin{document}
\begin{titlepage}
\begin{center}
April 2001\hfill    HIP-2001-09/TH

\vspace*{3cm}

{\large \bf g-2 of the muon in SUSY Models with Gauge
Multiplets in the Bulk of Extra-Dimensions}

\vskip .4in
Kari Enqvist$^{a,b,}$\footnote{kari.enqvist@helsinki.fi},
Emidio Gabrielli$^{a,}$\footnote{emidio.gabrielli@helsinki.fi},
Katri Huitu$^{a,}$\footnote{katri.huitu@helsinki.fi}\\[.15in]

{\em $^a$Helsinki Institute of Physics,
     and\\
     $^b$Physics Dept., University of Helsinki\\
     POB 64,\\
     00014 University of Helsinki, Finland}

\end{center}

\vskip .2in

\begin{abstract}

\end{abstract}
We analyze the supersymmetric contributions to the anomalous magnetic moment
of the muon ($\amu$) 
in the framework of pure and gaugino--assisted anomaly mediation 
models, and gaugino mediation models. In the last two models 
the gauge multiplets propagate in the higher dimensional bulk,
providing a natural mechanism for solving the problem 
of negative squared slepton masses present in the pure anomaly 
mediation models.
In the light of the new BNL results for $\amu$,
we found that the pure and gaugino-assisted
anomaly mediation models  are almost excluded by 
the BNL constraints at $2\sigma$ level when combined with 
CLEO constraints on \bsg at $90 \%$ of C.L.
In contrast, the gaugino mediation models provide 
extensive regions in the SUSY parameter space where both of these
constraints are satisfied.
\end{titlepage}

\newpage

\noindent
{}From the phenomenological point of view the breaking of supersymmetry 
(SUSY) is an essential issue in supersymmetric model building,
since it will determine the soft mass parameters, and therefore the 
mass pattern of the model.
After the detection of a few supersymmetric partners of ordinary
particles much can be deduced of the breaking mechanism.
Before direct detection, one can get indirect information from the
low energy signals.
In this respect, the measurement of the anomalous magnetic moment of the muon 
$(\amu)$ is a golden test of the SUSY contributions
beyond the Standard Model (SM).

Recently, the new measurement of $\amu$ at Brookhaven E821 experiment
\cite{E821}, has reported a $2.6 \sigma$ deviation from the SM prediction
\cite{Marciano}, 
\bea
a_{\mu}(\mbox{E821})-a_{\mu}(SM)= \left(43\pm 16\right) \times 10^{-10},
\label{BNLdev}
\eea
although this deviation has been criticized
\cite{Yndurain}, mainly due to the large theoretical uncertainties
affecting the hadronic contributions to the vacuum 
polarization\footnote{In reference 
\cite{newmarciano} it has been stressed that, although
some previous SM calculations agree with the experimental 
result for $a_\mu$ \cite{SMagree}, they are not 
on a equal footing or up--to--date in comparison with the 
analysis of Davier and H\"ocker in \cite{Marciano}.}.
However, the possibility that new physics is responsible for
such a deviation, has motivated an intense theoretical activity 
especially in the framework of supersymmetric theories \cite{g2_new}.
Indeed, if this deviation has a supersymmetric origin, 
the lightest SUSY particles should 
be in the expected discovery range of the Large Hadronic $pp$ 
Collider (LHC) at CERN and even of the Fermilab 2 TeV $p\bar{p}$ collider.

The aim of the present paper is to analyze, in the light of the
BNL deviation,
the predictions for $\amu$ in a special class of supersymmetric models,
where the SUSY breaking is realized with gauge multiplets 
propagating in the bulk of extra dimensions.
In recent years several interesting aspects of supersymmetry breaking
have emerged.
Especially the branes, which are typical in models with extra
dimensions, have been found to fit naturally with the idea of breaking  
supersymmetry in a hidden sector.

In this framework, the anomaly mediated breaking (AMSB) 
\cite{anomaly} is particularly
attractive, since this contribution to the breaking will always 
be present.
Unfortunately this mechanism has a well known flaw: squares of the
slepton masses become negative.
Because of the other virtues of the model, several methods have
been proposed in order to remove this problem \cite{anomalyfix}.
Assuming gauge multiplets in the higher dimensional bulk, it was
found in \cite{Kaplan} that at one loop the squared slepton masses 
obtain  contributions, which would be of the correct size for solving 
the slepton mass problem.

Another breaking mechanism connected with the extra dimensions is
the so-called gaugino mediation ($\tilde g$MSB) \cite{gaugino}, in 
which the gauge superfields propagate in the bulk in addition to 
gravity, and couple at tree-level to a singlet at SUSY breaking brane.
This mechanism does not have any problems with experiments,
but it has some arbitrariness in the scale of the compactification
radius.

We will work in the framework proposed by Randall and Sundrum in 
\cite{anomaly},
where two branes are located in one extra dimension.
One of the branes contains the hidden sector, while the other
brane contains the ordinary matter.
Gravity is in the bulk.
Here we will assume that also the gauge supermultiplets reside in
the bulk, following \cite{Kaplan, gaugino}.

If there is no coupling of the gauge fields to the fields in the
hidden sector, the model is the one presented in \cite{Kaplan},
in which the squared scalar masses, in addition to the usual
contributions from superconformal anomaly, receive contributions at 
one loop due to the corrections to the wave function renormalization 
of gauginos.

In pure anomaly mediation the soft supersymmetry breaking parameters
are obtained as \cite{anomaly}
\bea
M_a&=&\frac{\beta_{g_a}}{g_a}m_{3/2},\nonumber\\
\tilde m_i^2&=&-\frac 14 \left(\frac{\partial\gamma_i}{\partial g}
\beta_g +\frac{\partial\gamma_i}{\partial Y}\beta_Y \right) m_{3/2}^2,
\label{pureanom}
\\
A_Y&=&-\frac{\beta_Y}{Y} m_{3/2},\nonumber
\eea
where $\beta_i$ and $\gamma_i$ are the beta functions and anomalous
couplings, respectively.
The explicit expressions can be found in \cite{gherghetta}.
It is obvious that the slepton mass squared
containing positive $SU(2)$ $\beta$-functions are negative.
The extra contribution $\Delta \tilde m^2_i$ 
to $\tilde m^2_i$ in (\ref{pureanom}), received in the 
framework of \cite{Kaplan}, is given by
\bea
\Delta \tilde m^2_i =2\zeta(3)\Gamma(4)\sum_a C_a(i)\frac{g^2}{16\pi^2}\frac 1{(M_*L)^2}
m_{3/2}^2 \equiv \eta\sum_a C_a(i) \frac{g^2}{16 \pi^2} m_{3/2}^2,
\label{gassistedmass}
\eea
where $\sum_c C_a(i)$ is the weighted sum over the
quadratic Casimir for the $i$ matter scalar representations, 
tabulated in \cite{Kaplan}, and $g$ is the gauge coupling at unification scale.
The Casimir factors will give to the scalar masses a nonuniversal 
contribution, which is the major difference between the 
gaugino-assisted and minimal anomaly mediation models.
In the latter, $\Delta \tilde m^2_i$ is assumed universal.
{}From the requirement of small flavour violating operators
\cite{Kaplan}, the volume factor $M_*L\gsim 16$.
Thus the numerical value of $(M_*L)^{-2}\sim g^2/(16 \pi^2)$.
On the other hand the scalar masses are two-loop suppressed in
pure anomaly mediation, in
contrast to the one-loop suppression in Eq. (\ref{gassistedmass}).
Accidentally, therefore, the extra contribution to the scalar masses is 
of the correct size to solve the slepton mass problem in AMSB.
$\eta$ will be used as the parameter in our numerical
calculations.

If there is a direct coupling of gauge fields to the SUSY breaking
brane, with a singlet which receives a VEV, the gauginos get a SUSY
breaking mass \cite{gaugino}.
Minimally the gaugino mediation model has three parameters,
namely the Higgs mixing parameter $\mu$, the common gaugino mass
$M_{1/2}$ and the compactification scale $M_c$.
Following Schmaltz and Skiba in \cite{gaugino}, we assume that at 
the GUT scale the soft breaking
$A$ parameters, as well as the soft scalar masses, vanish
and the compactification scale is in the range 
$M_{GUT}\lsim M_c\lsim M_{Planck}/10$.
However, since we are interested in analyzing a more general scenario than in 
\cite{gaugino}, we will take $\tgb$  as a free parameter.
This is effected by relaxing the condition 
of vanishing soft $B$ parameter at GUT scale assumed in \cite{gaugino}.
The main reason for our choice is that, as shown in the following, 
the SUSY contribution to $\amu$ is very sensitive to $\tgb$.

In gaugino mediation models the gaugino mass is proportional to the
$F$ component of a chiral superfield on the SUSY breaking brane.
To be more specific, with a singlet $S$, the gaugino masses are
generated by
\bea
\int \, d^2\theta \frac {S}{M_*^2} W^\alpha W_\alpha \delta (y-L),
\eea
where $y$ is the extra dimensional coordinate, $L$ is the place of the
SUSY breaking brane, and the resulting gaugino mass is
$M_{1/2}\propto F_S/(M_*^2L)$.
If the VEV $F_S$ does not exist, the gaugino masses could be generated
by some higher dimensional operators, containing charged fields $\Sigma$
in extra dimensions,
\bea
\int \,d^4\theta \frac{\Sigma^\dagger\Sigma}{M_*^2} W^\alpha
W_\alpha \delta (y-L),
\eea
giving a contribution of order $F_\Sigma^2/(M_*^4 L)$.
Assuming that $F_S$ and $F_\Sigma$ are of the same order, the
anomaly induced contribution is negligible.  
However, if the magnitudes of $F_S$ and $F_\Sigma$ differ 
considerably, or one of them does not exist, it is of interest to study the 
situation in which one of these is dominant.

Now we analyze the SUSY contributions to $\amu$ in the framework 
of the anomaly and gaugino mediation models.
The minimal supersymmetric standard model (MSSM) contributes to $\amu$ 
mainly via magnetic--dipole penguin diagrams, with an exchange of a chargino
or a neutralino in the loop. 
As is well known, the $\amu$ is enhanced by 
$\tgb$, which is defined as  $\tgb=\frac{<H_U>}{<H_D>}$.
In the large $\tgb$ limit, the $\asusy$ is dominated by the chargino 
diagram which is given by \cite{g2_gs}
\be
a_{\mu}^{{\rm SUSY}}\approx \frac{3\alpha_2}{4\pi}\tgb 
\frac{m_{\mu}^2 \mu M_2}{m^4_{\tilde{\nu}}} F(x_{M_2},x_{\mu})\ ,
\label{g-2}
\ee
where $M_2$ is the weak gaugino mass, $m_{\tilde{\nu}}$ is the 
sneutrino mass and $F$ is a loop function defined as
\be
F(x,y)=\frac{f(x)-f(y)}{x-y}\
,~~~~~f(x)=\frac{3-4x+x^2+2\log(x)}{3(1-x)^3}\ ,
\label{loopF}
\ee
where $x_{M_2}=M_2^2/m^2_{\tilde{\nu}},~~x_{\mu}=\mu^2/m^2_{\tilde{\nu}}$.

From Eq.(\ref{g-2}) we see that the free parameters entering 
in the $\asusy$ formula at large $\tgb$ 
are the $\mu$ parameter, the sneutrino mass 
$m_{\tilde{\nu}}$, the weak gaugino mass $M_2$, and $sign(\mu M_2)$.
In addition, the sub-dominant neutralino diagram contribution depends
also on the smuon  and neutralino masses.
However, in our analysis we have used the exact
one-loop expression for $\asusy$ which could be found, 
for instance, in Ref. \cite{g2_rev}.

The ingredients used in our analysis can be summarized as follows.
The SUSY masses and parameters at electroweak scale are obtained 
from their boundary conditions at GUT scale, by
solving the one-loop renormalization group equations of MSSM \cite{Ibanez}.
Moreover, we compute the $\mu$ parameter by requiring 
the condition of the correct electroweak symmetry breaking.
In the computation of gaugino masses at low energy, 
we have included the next--to--leading order corrections in $\alpha_s$ and
the top Yukawa coupling \cite{gherghetta}. 
Indeed these corrections may become relevant
for the anomaly--mediation scenarios, where the triplet of Winos 
$(\tilde{W}^{\pm}, \tilde{W}^{0})$ is nearly degenerate.
Also in the calculation of the
lightest Higgs mass $m_h$ we have included the leading one-loop 
radiative corrections \cite{haber}.

In order to avoid tachyons and vacuum instabilities, we require
that all the scalar masses at the GUT scale are positive. Besides,
the present lower bounds on the SUSY particle masses 
from accelerator experiments have been imposed.
In particular for the lightest chargino and Higgs masses, 
we use $m_{\chi^{\pm}} > 83 $ GeV
and $m_{h} > 114 $ GeV, respectively. Note that our lower bound on
chargino mass is suitable for the class of
models considered here, see Ref. \cite{LEPtalk} for further details.

In addition, we have imposed the CLEO bounds on the decay
$B\to X_s \gamma$ at 90\% C.L. \cite{CLEO}
\bea
2.0\times 10^{-4}<{\rm BR}(B\to X_s\gamma)<4.5\times 10^{-4}.
\label{cleo}
\eea
In MSSM, the SUSY contribution to \bsg comes from the flavour-changing
magnetic--dipole penguin diagrams with exchanges of charged Higgs,
chargino, neutralino, and gluino \cite{BBMR}.
When the scalar soft--breaking terms at GUT scale are universal in flavour, 
such as for instance in the models considered here, the main contributions come
from the diagrams with an exchange of charged Higgs and charginos.
In our analysis we used the
parametrization of \cite{g2_gs} for the SUSY contribution to the 
$B \to X_s \gamma $ decay, which includes the next--to--leading order 
QCD corrections for the SM contribution,
and the leading--order SUSY contributions to the relevant Wilson 
coefficients.\footnote{We have used in our analysis
the central value for the SM branching ratio 
$BR(B \to X_s \gamma )=3.28\times 10^{-4}$. Note that in Ref. \cite{g2_gs}
the corresponding one was a bit higher, due to 
some (percent--level) non--pertubative corrections included in that 
evaluation.}

As known, the \bsg constraint plays a crucial role in the $\asusy$ 
analysis, in particular at large $\tgb$.
The reason is that in MSSM the chargino contribution to 
the \bsg amplitude is enhanced by $\tgb$, and at large
$\tgb$ it might become comparable in magnitude with the charged Higgs one, 
when chargino masses are quite light \cite{BBMR}.
Moreover the charged Higgs contribution has always a positive 
interference with the SM amplitude, 
while the sign of the chargino contribution crucially 
depends on $sign(\mu)$ (when the sign of $M_2$ is fixed).\footnote{
In some models the sign of $M_2$ is also a free parameter and therefore
the physical sign in this case is given by $sign(\mu M_2)$.}
Therefore $sign(\mu)$ plays a decisive role in selecting the regions 
where the chargino and charged Higgs have
constructive or distructive interferences,
disfavoured or favoured by the CLEO constraints, respectively.

An important consequence of the BNL constraints on MSSM is that
the value of $sign(\mu)$ for which $\asusy$ becomes negative is not allowed.
When the BNL constraints on $\amu$
are combined with the CLEO ones on \bsg, 
then there is no more freedom in choosing
$sign(\mu)$ in order to favour these last ones. This effect
might strongly reduce the allowed SUSY parameter space.  As shown in
the following this will be the case for the anomaly mediation models.

\begin{figure}[tpb]
\dofigs{3.1in}{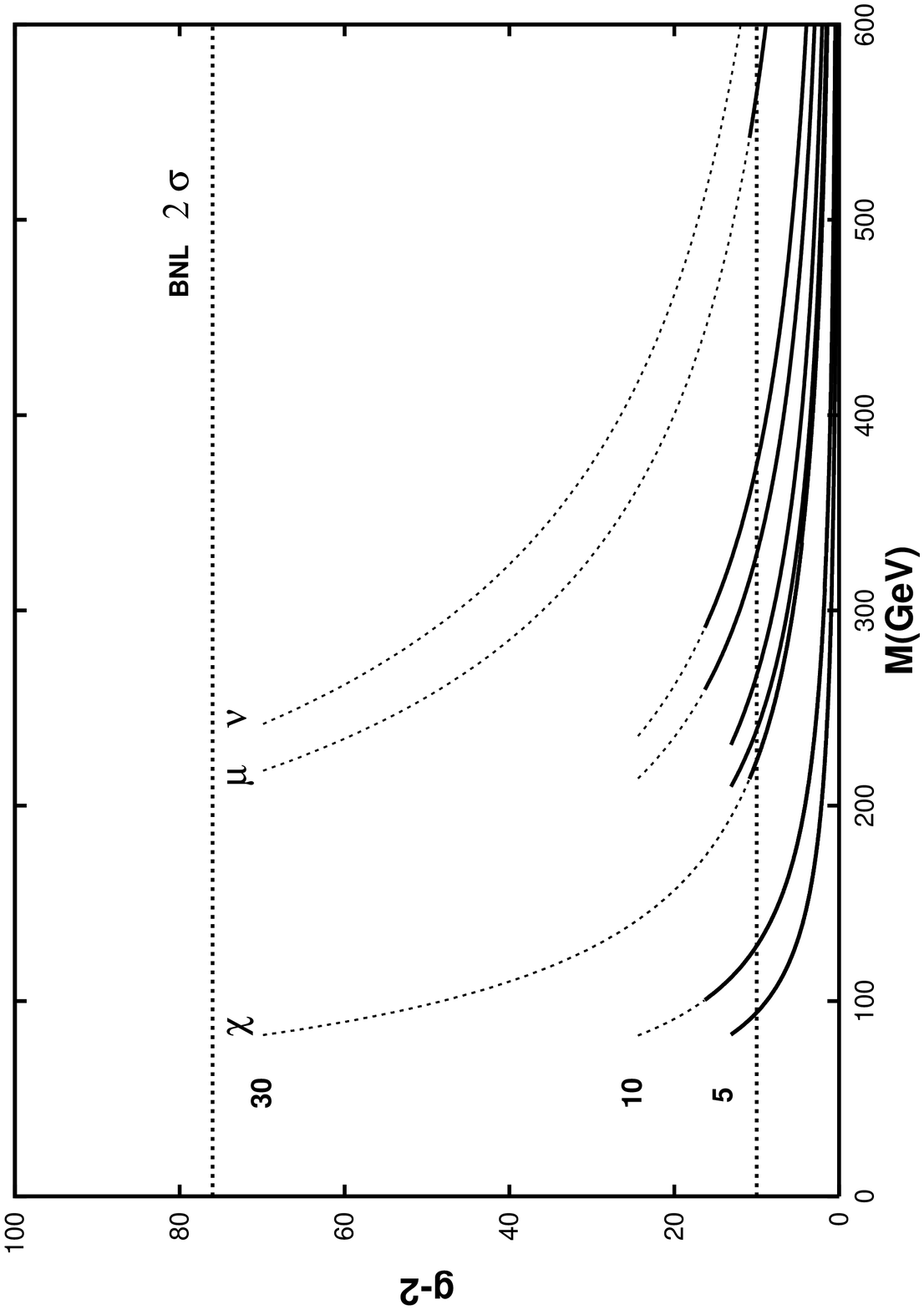}{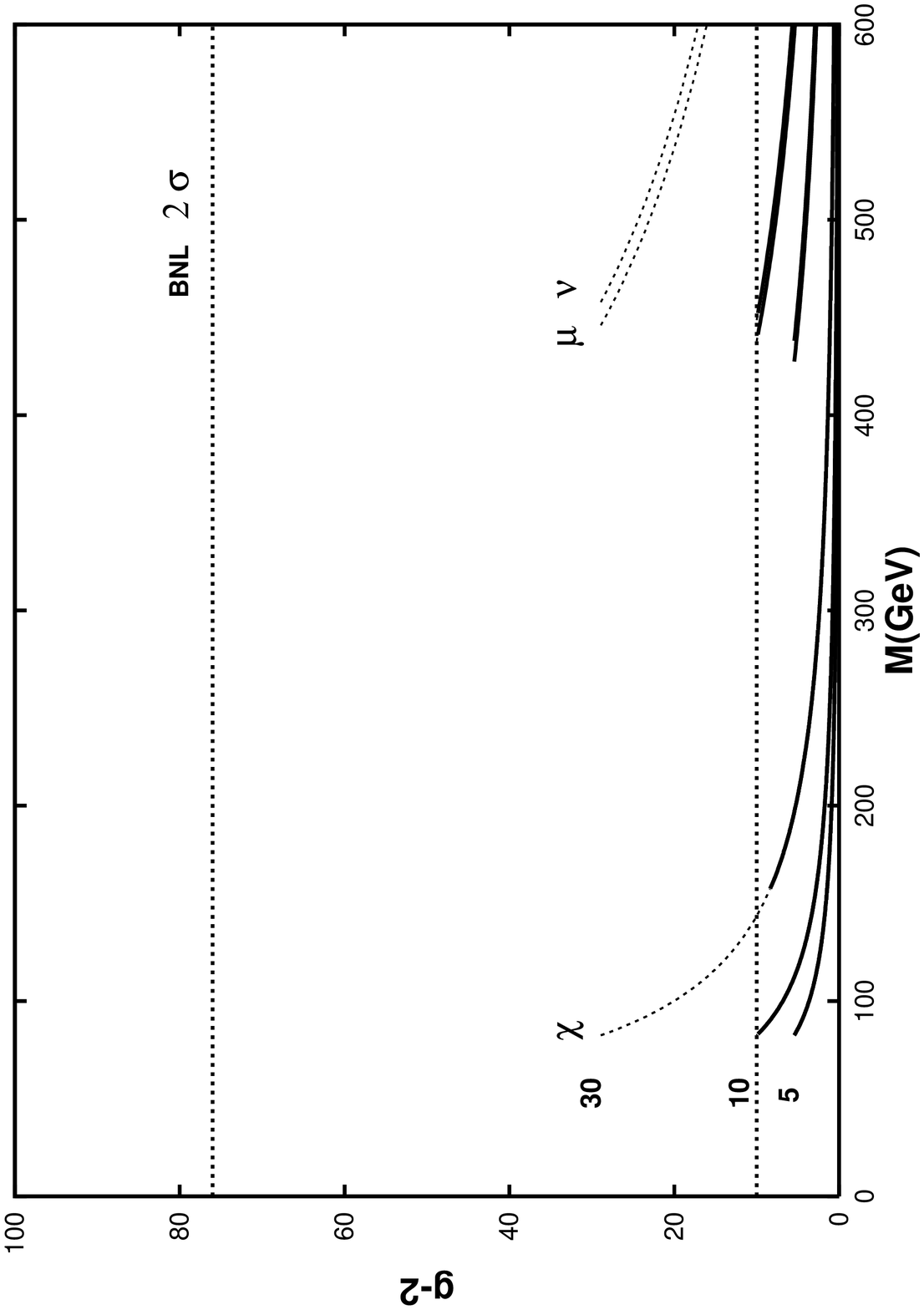}
\caption{{\small In the minimal anomaly mediation model, 
$\asusy$ (in units of $10^{-10}$) versus the lightest 
chargino ($\chi$), smuon ($\mu$), and sneutrino ($\nu$) 
masses, for 
three values of $\tgb=5,10,30$. The left and right plots 
correspond to  $\eta=0.03$ and $\eta=0.1$ respectively, where $\eta$
is defined as in Eq.(\ref{gassistedmass}) with $\sum_a C_a(i)=1$.
The area inside the two horizontal lines stands for the BNL deviation 
at $2\sigma$ level. The dashed lines represent regions excluded by the 
\bsg constraints.}}
\label{fig1}
\end{figure}

\begin{figure}[tpb]
\dofigs{3.1in}{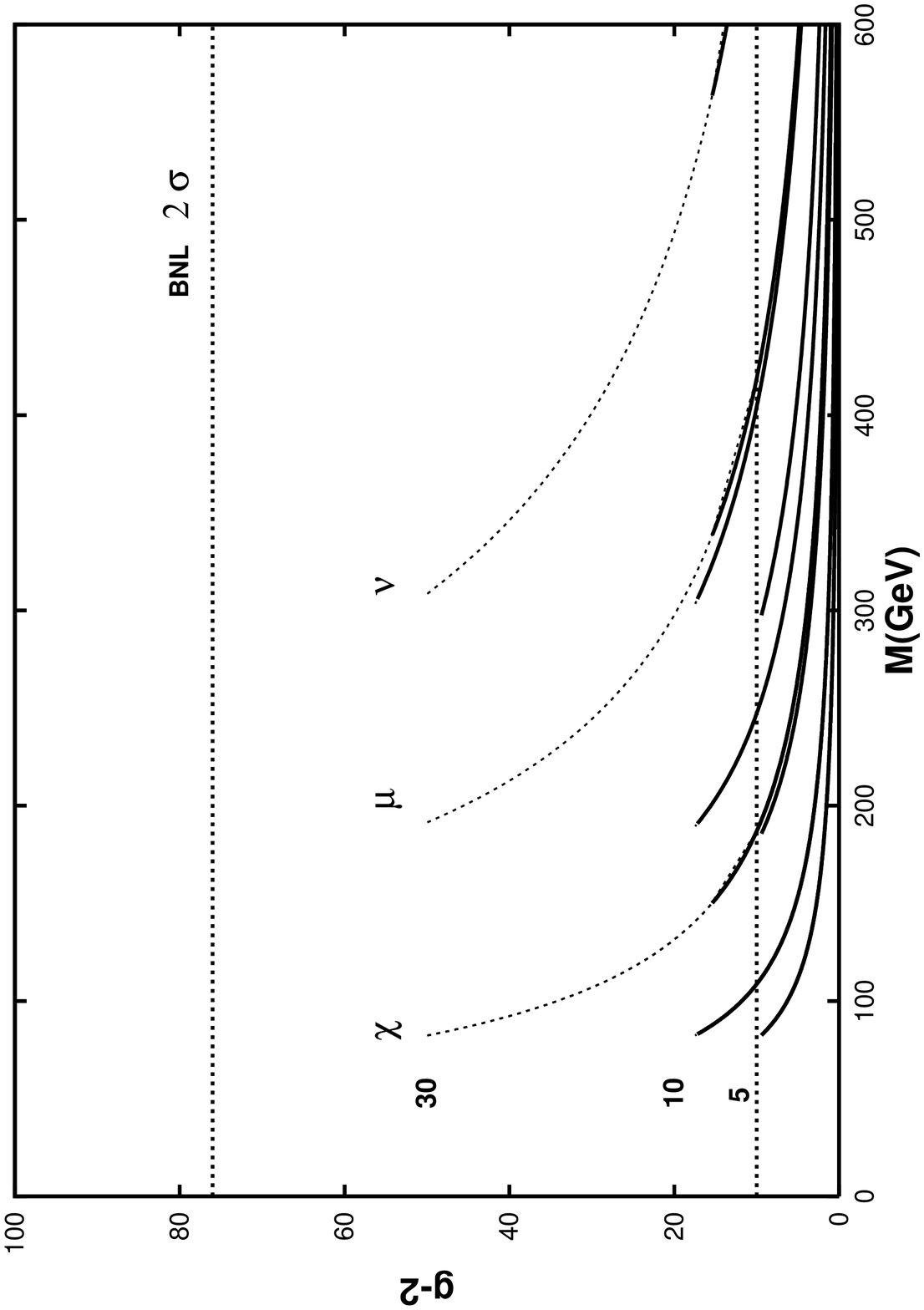}{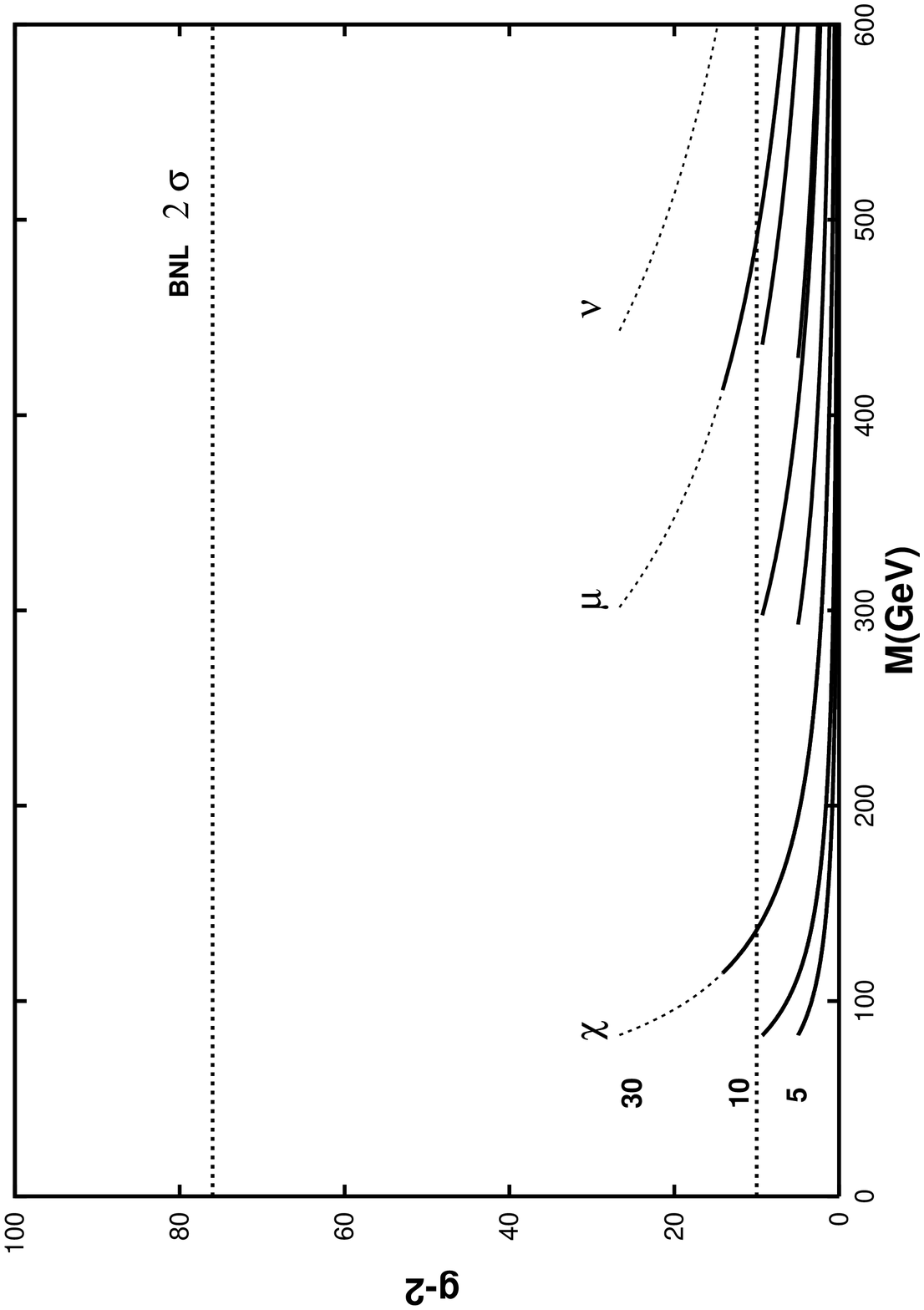}
\caption{{\small As in Fig. (\ref{fig1}), but for the 
gaugino--assisted anomaly mediation model, where $\eta$ is defined in 
Eq.(\ref{gassistedmass}). The left and right plots 
correspond to $\eta=0.05$ and $\eta=0.1$ respectively.}}
\label{fig2}
\end{figure}

\begin{figure}[tpb]
\dofigs{3.1in}{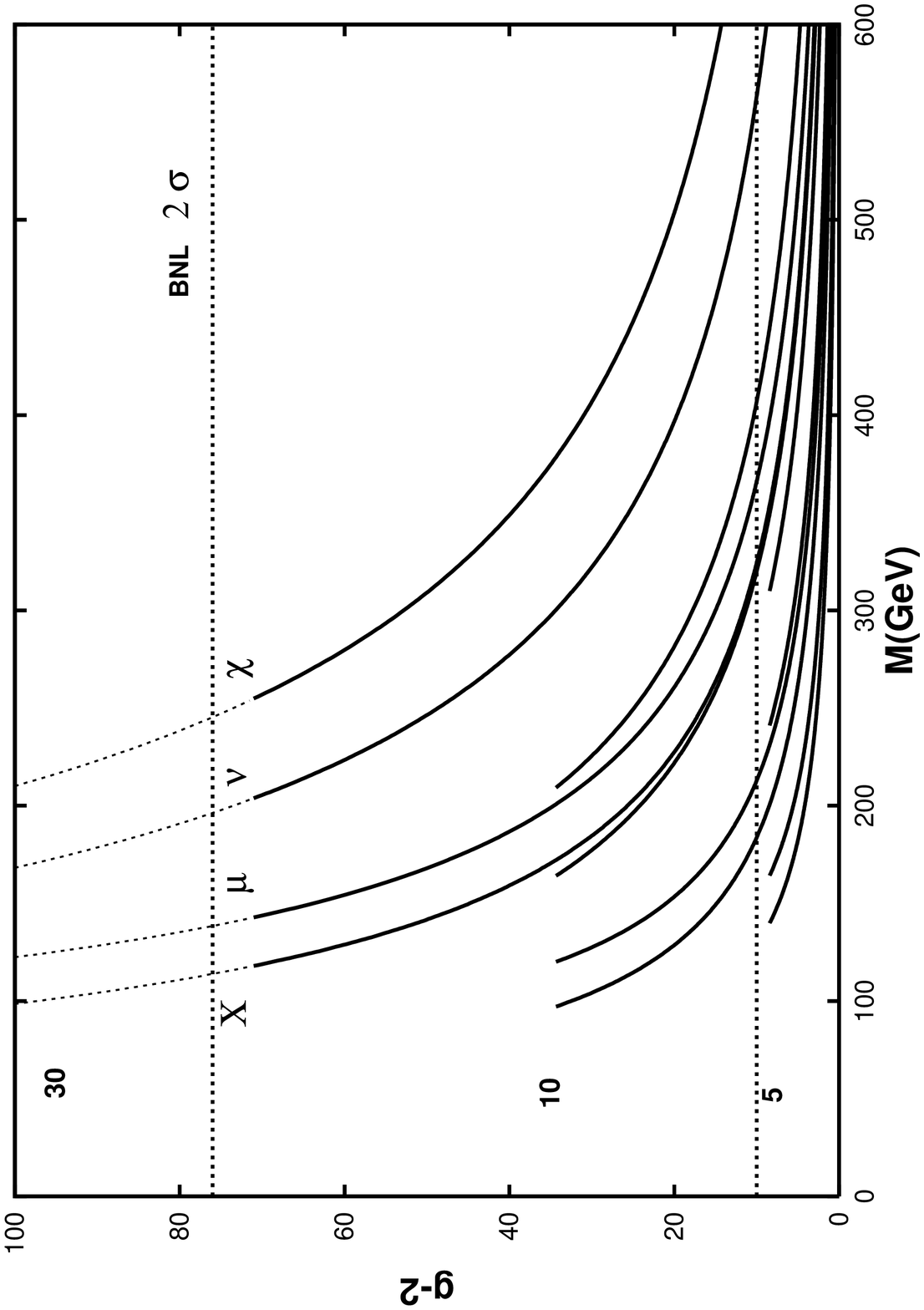}{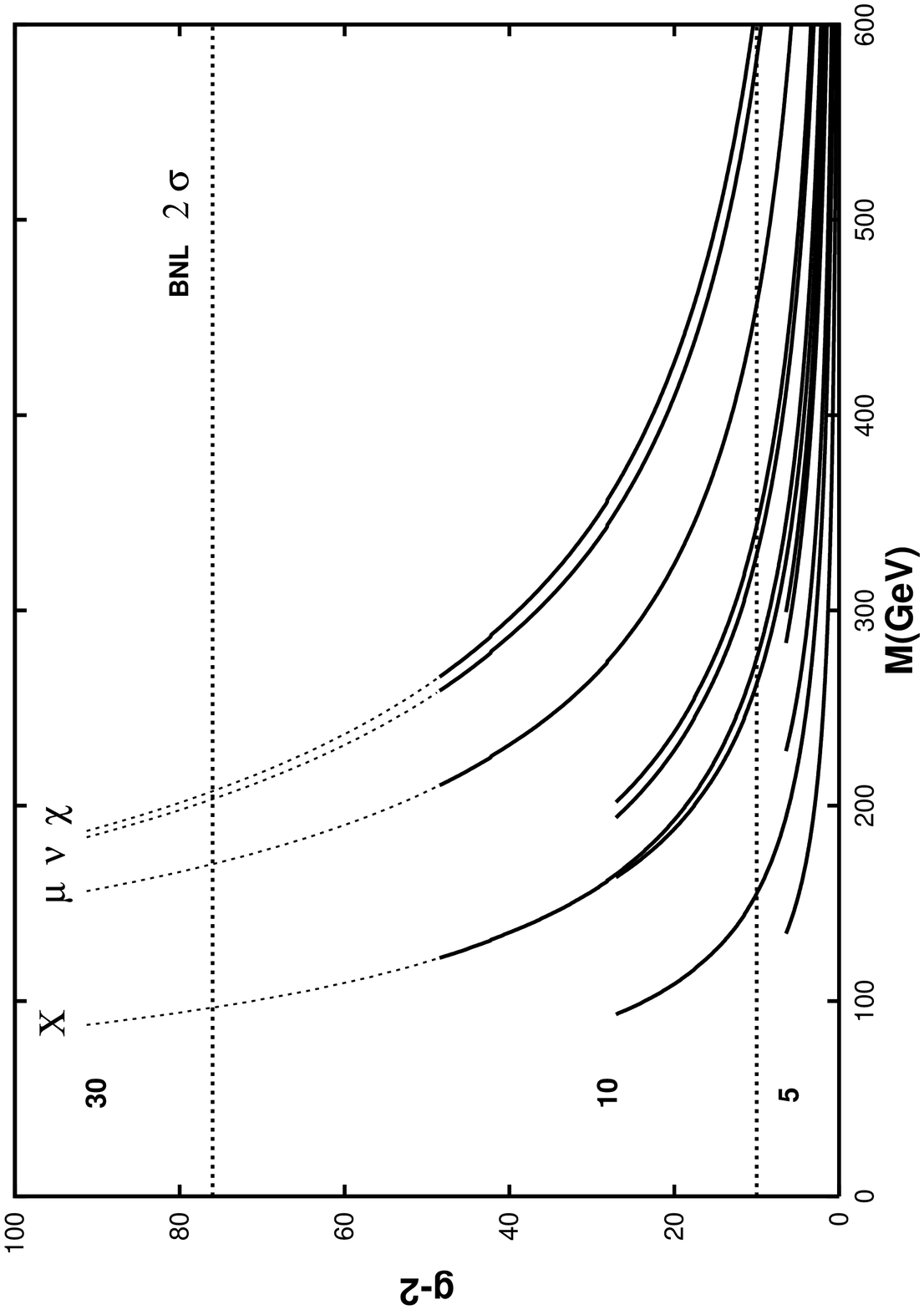}
\caption{{\small As in Fig. (\ref{fig1}), but for
the minimal gaugino mediation model with $SU(5)$ unified group.
The label on the curves stand for the lightest 
neutralino ($X$), chargino ($\chi$), smuon ($\mu$), and sneutrino ($\nu$).
The left and right plots correspond to $t_c=1$ and $t_c=4$ respectively, where
$t_c={\rm Log} (M_c/M_{GUT})$.}}
\label{fig3}
\end{figure}

\begin{figure}[tpb]
\dofigs{3.1in}{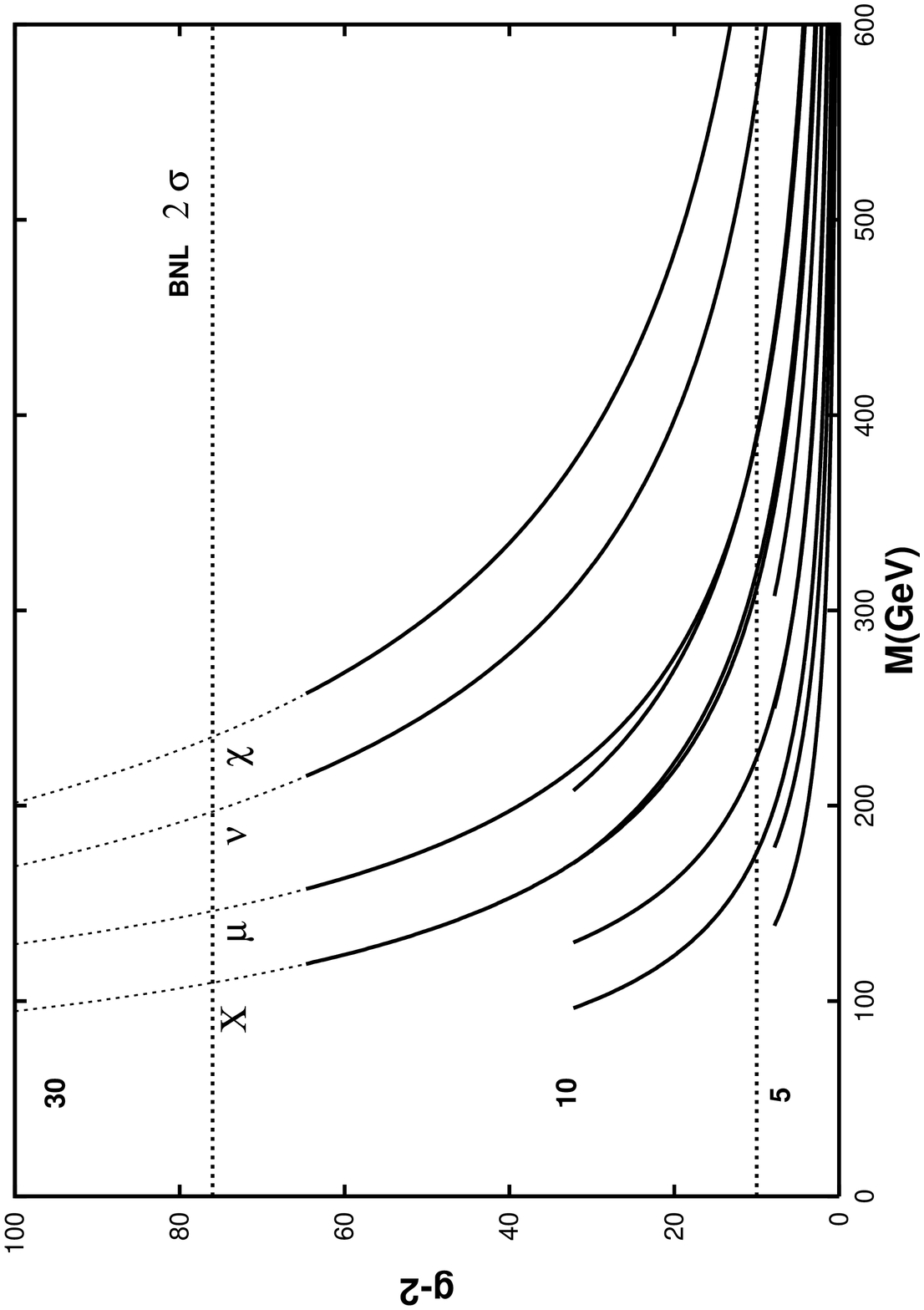}{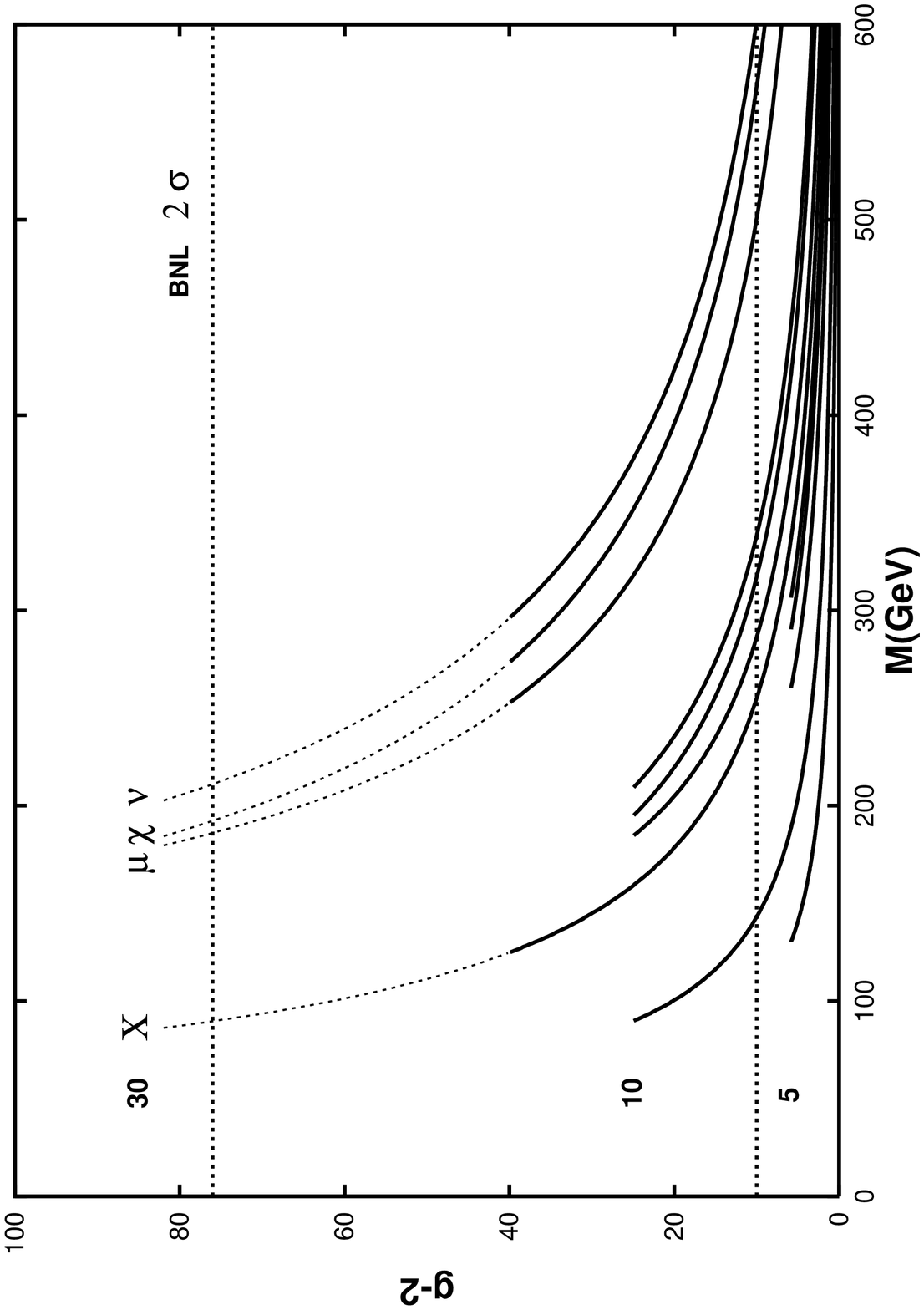}
\caption{{\small As in Fig. (\ref{fig3}), but with $SO(10)$ unified group.
The left and right plots correspond to $t_c=1$ and $t_c=4$ respectively.}}
\label{fig4}
\end{figure}

Our numerical results are shown in Figs. (\ref{fig1}-\ref{fig2}), 
for the anomaly mediation and 
gaugino assisted anomaly--mediation respectively, and in
Figs. (\ref{fig3}-\ref{fig4})
for the minimal gaugino mediation models.
In all these figures the region inside the two horizontal lines indicate the
BNL deviation in Eq.(\ref{BNLdev}) at $2\sigma$ level.
The $\asusy$ has been plotted versus 
the relevant low energy spectrum, such as the 
lightest chargino, neutralino, smuon, and sneutrino masses.
We show the curves only for three
representative values of $\tgb=5,10,30$, where the dashed lines 
always indicate the regions excluded by the \bsg constraints.

Let us first discuss the results in the anomaly mediation scenarios.
In Figs. (\ref{fig1}-\ref{fig2}), the curves corresponding to the 
lightest neutralino or chargino roughly coincide since
in these models the lightest chargino and neutralino masses 
are almost degenerate, so we show only the chargino ones.
In the minimal anomaly mediation scenario we parametrize $\tilde m^2$ as 
appearing in Eq.(\ref{gassistedmass}), but with $\sum_a C_a(i)=1$.

The left and right plots in Fig. (\ref{fig1}) correspond respectively to 
$\eta=0.03$ and $\eta=0.1$.
In particular, the value of $\eta=0.03$ corresponds to the allowed minimum 
for $\tilde m$ (at fixed $m_{3/2}$) for which no tachyons
in the slepton sector will appear. The larger the $\eta$ is, 
the larger the SUSY scalars become at fixed chargino or neutralino masses.
From the results of Fig. (\ref{fig1}) one can see that already at $\tgb=30$
the combined effect of BNL and CLEO bounds exclude
the minimal anomaly mediation scenario, while
at intermediate $\tgb$ regions, such as $\tgb=10$, 
this model is barely allowed.
We note that the 
lower bounds on the stau and lightest Higgs mass prevent
the chargino mass going below $200$ GeV for $\tgb=30$.

Clearly, when the scalar masses are increasead, $\eta=0.1$, 
the regions at moderate $\tgb$ are also excluded due to the SUSY 
decoupling, as can be seen comparing 
the left and right plots in Fig. (\ref{fig1}).
Then the main conclusion from these results is that the anomaly 
mediation models
seem to be disfavoured in explaining the $2\sigma$ BNL deviation, while
preserving the CLEO bounds in Eq. (\ref{cleo}). 
These results are also in agreement with the corresponding ones in 
\cite{fengmoroi} and in the second reference of \cite{g2_new}.

In Fig. (\ref{fig2}) we show the corresponding results in
the gaugino--assisted anomaly mediation model. This model
differs from the minimal anomaly--mediation one only by the fact that here the
extra contributions to the scalar masses in Eq. (\ref{gassistedmass}) are not
universal for all the scalars. As in Fig. (\ref{fig1}),
$\eta=0.05$ correspond here to the minimal allowed value for $\tilde{m}$, 
(at fixed $m_{3/2}$). Note that the minimum value for $\eta$ 
differs from the previous scenario due to the presence of
the Casimir factors in Eq.(\ref{gassistedmass}).
Also in this scenario we get similar 
results with respect to the minimal anomaly mediation one. From the results 
in Fig. (\ref{fig2}) we see that the non-universality 
in the scalar sector does not 
help very much in getting allowed extensive regions where the CLEO 
and BNL constraints are satisfied.
However, we see that, with respect to the minimal anomaly mediation scenario, 
also in this case the intermediate $\tgb$ regions, 
such as $\tgb=10$, are barely favoured in satisfying these constraints.

In Fig. (\ref{fig3}) we show the results for the minimal gaugino 
mediation model with SU(5) unified group.
In this scenario the free parameters 
are the common gaugino mass $M$,
the parameter $t_c={\rm Log}(M_{c}/M_{GUT})$, 
where $M_c$ is the compactification scale, $\tgb$, and $sign(\mu)$.
Since the scalar masses $\tilde m$ and trilinear couplings $A$ are
radiatively generated through the renormalization group running
from the compactification scale down to the GUT scale, 
they will be proportional respectively to 
$\tilde m^2 \propto M^2~\alpha~t_c$, and 
$A \propto M~\alpha~t_c$ times averaged Casimir factors.
For the exact expressions we have
used the results of the last reference in \cite{gaugino}.
In Fig.[\ref{fig3}] we have shown two representative examples 
with $t_c=1$ and $t_c=4$ for the left and right plot respectively, 
and for $\tgb=5,10,30$.
For dashed and continuous curves we use the same convention
adopted in Figs. (\ref{fig1}-\ref{fig2}).
From Fig. (\ref{fig3}) we see that this class of models are
favoured in obtaining extensive regions in the parameter space where
$\asusy$ is within the $2\sigma$ BNL deviation, while respecting the 
CLEO constraints, even at intermediate $\tgb$ regions ($\tgb > 10$). 
Moreover the regions with $\tgb <5$ seem to be excluded 
by the lower bounds of the $2\sigma$ BNL constraints.
Note that, with respect to the anomaly mediation models, the 
hierarchy of the SUSY particles masses plotted in Fig. (\ref{fig3})
is changed. Now the neutralino is the lightest one, 
followed in order by the smuon, sneutrino, and chargino ones.
Increasing the value of compactification scale, 
the lightest chargino and sneutrino masses become more degenerate, 
as shown in the right plot of Fig. (\ref{fig3}).

In Fig. (\ref{fig4}), the results for the minimal gaugino 
mediation model with the SO(10) unified group are shown.
The difference with the corrseponding results
in Fig. (\ref{fig3}) lies in the different boundary
conditions at GUT scale \cite{gaugino}.
As in Fig. (\ref{fig4}) we see that also for this model there are
extensive regions in the parameter space where both BNL and CLEO
constraints are satisfied. Besides, as in the case of SU(5), the regions with
$\tgb <5 $ are excluded by the lower bound of the $2\sigma$ BNL constraints.
Note that when $t_c=4$ the sneutrino becomes 
heavier than the lightest chargino one, for the region of masses and $\tgb$
considered in Fig. (\ref{fig4}).

Thus the remarkable aspect of the BNL deviation is that 
at $2\sigma$ level one can obtain upper 
limits on the lightest chargino ($m_{\chi^{\pm}}$),  
neutralino ($m_{\chi^{0}}$), smuon ($\tilde{m}_{\mu}$), 
and sneutrino ($\tilde{m}_{\nu_{\mu}}$) masses.
The correspondig (conservative) upper limits in the minimal gaugino mediation
can be summarized as follows.
In the case of SU(5) we obtain, for $t_c=1~ (4)$ and $\tgb \le 30$, the
following results
$m_{\chi^{0}} \le 320~ (280)$ GeV,~~$m_{\chi^{\pm}} \le 720~ (610)$ GeV,~~
$\tilde{m}_{\mu} \le 370~ (460)$ GeV, and  
$\tilde{m}_{\nu_{\mu}} \le 560~ (580)$ GeV, while the corresponding ones
for SO(10) are given by
$m_{\chi^{0}} \le 310~ (250)$ GeV,~~$m_{\chi^{\pm}} \le 690~ (570)$ GeV,~~
$\tilde{m}_{\mu} \le 390~ (500)$ GeV, and  
$\tilde{m}_{\nu_{\mu}} \le 560~ (600)$ GeV.
As we can see, these upper limits are not very sensitive to the different
structure of SU(5) and SO(10) unified gauge group.

In this framework we have also analyzed the implications 
of the gaugino mediation scenarios 
for current dark matter detectors, DAMA \cite{dama} and CDMS \cite{cdms}.
Indeed it is known that the lightest neutralino is the natural candidate 
for dark matter in SUSY models with conserved R parity.
These detectors are sensitive to the neutralino--nucleon cross section 
($\sigma_{\chi}$) in the range of $10^{-6}-10^{-5}$ pb. 
Recently in the framework of MSSM 
with universal boundary conditions 
it was pointed out that the large $\tgb$ regime allows $\sigma_{\chi}$
to reach the above range \cite{darkmatter}. 
In addition, with non--universal soft scalar 
masses it is possible to obtain large cross sections even for moderate 
$\tgb$ regions \cite{darknouniv}. We checked that
in the minimal gaugino mediation scenario, where non--universal soft
masses are naturally generated at GUT scale \cite{gaugino},
the neutralino--nucleon cross section is unfortunately not enhanced.
In particular we found that in both $SU(5)$ and $SO(10)$ models
and in the parameter region where $\asusy$ is positive,
$\sigma_{\chi}$ is always below $10^{-7}$ pb for $\tgb \leq 30$ 
and $1<t_c<4$.

In conclusion, assuming that the complete effect of the BNL deviation is
due to supersymmetry, we find that the minimal and gaugino-assisted
anomaly mediation models,  seem to be both disfavoured by 
the BNL constraints at $2\sigma$ level when combined with the
CLEO constraints on \bsg at $90 \%$ of C.L.
On the contrary, in the minimal gaugino mediation models 
we found extensive regions in the SUSY parameter space, where these
constraints are satisfied. 
Thus for these models, we have obtained conservative upper bounds 
on the relevant SUSY spectrum. These are of the order of
300 GeV and 700 GeV for the lightest neutralino and chargino respectively, 
and of the order of 500 GeV and 600 GeV for the lightest smuon and sneutrino
respectively, when $\tgb \le 30$ and the compactification scale is in the range
$1\le t_c \le 4$.

\vspace*{1cm}
\noindent
{\bf Acknowledgements}

\vspace*{0.5cm}
\noindent
This work was partially supported by the Academy of Finland
(project nos. 35224, 48787, and 163394).

\end{document}